\documentstyle[11pt,amssymb]{article}

\def\dalemb#1#2{{\vbox{\hrule height .#2pt
        \hbox{\vrule width.#2pt height#1pt \kern#1pt
                \vrule width.#2pt}
        \hrule height.#2pt}}}

\def\0{{\sst{(0)}}}
\def\1{{\sst{(1)}}}
\def\2{{\sst{(2)}}}
\def\3{{\sst{(3)}}}
\def\4{{\sst{(4)}}}
\def\5{{\sst{(5)}}}
\def\6{{\sst{(6)}}}
\def\7{{\sst{(7)}}}
\def\8{{\sst{(8)}}}

\def\Z{\rlap{\sf Z}\mkern3mu{\sf Z}}
\def\R{\rlap{\rm I}\mkern3mu{\rm R}}

\def\td{\tilde}
\def\wtd{\widetilde}

\let\a=\alpha

\def\nn{\nonumber} \def\bd{\begin{document}} \def\ed{\end{document}}
\def\ds{\documentstyle} \let\fr=\frac \let\bl=\bigl \let\br=\bigr
\let\Br=\Bigr \let\Bl=\Bigl 
\let\bm=\bibitem
\let\na=\nabla
\let\pa=\partial \let\ov=\overline 
\newcommand{\be}{\begin{equation}} 
\newcommand{\ee}{\end{equation}} 
\def\ba{\begin{array}}
\def\ea{\end{array}}
\def\ft#1#2{{\textstyle{{\scriptstyle #1}\over {\scriptstyle #2}}}}
\def\fft#1#2{{#1 \over #2}}
\def\del{\partial}
\def\sst#1{{\scriptscriptstyle #1}}
\def\oneone{\rlap 1\mkern4mu{\rm l}}
\def\ie{{\it i.e.\ }}
\def\via{{\it via}}
\def\semi{{\ltimes}}
\def\str{{\rm str}}
\def\jm{{\rm j}}
\def\im{{\rm i}}
\def\bOmega{{{\bar\Omega}}}
\def\Qn{{{Q_{\sst{\rm N}}}}}

\def\mapright#1{\smash{\mathop{-\!\!\!-\!\!\!-\!\!\!-\!\!\!-\!\!\!
             \longrightarrow}\limits^{#1}}}
\def\maprightt#1#2{\smash{\mathop{-\!\!\!-\!\!\!-\!\!\!-\!\!\!-\!\!\!
             \longrightarrow}\limits^{#1}_{#2}}}

\newcommand{\ho}[1]{$\, ^{#1}$}
\newcommand{\hoch}[1]{$\, ^{#1}$}
\newcommand{\bea}{\begin{eqnarray}} 
\newcommand{\eea}{\end{eqnarray}} 
\newcommand{\ra}{\rightarrow}
\newcommand{\lra}{\longrightarrow}
\newcommand{\Lra}{\Leftrightarrow}
\newcommand{\ap}{\alpha^\prime}
\newcommand{\bp}{\tilde \beta^\prime}
\newcommand{\tr}{{\rm tr} }
\newcommand{\Tr}{{\rm Tr} }

\newcommand{\NP}{Nucl. Phys. }
\newcommand{\tamphys}{\it Center for Theoretical Physics\\
Texas A\&M University, College Station, Texas 77843}
\newcommand{\ens}{\it Laboratoire de Physique Th\'eorique de l'\'Ecole
Normale Sup\'erieure\hoch{2,3}\\
24 Rue Lhomond - 75231 Paris CEDEX 05}
\newcommand{\upenn}{\it Department of Physics and Astronomy\\
University of Pennsylvania, Philadelphia, Pennsylvania 19104}

\newcommand{\auth}{M. Cveti\v{c}\hoch{\dagger1}, 
H. L\"u\hoch{\dagger1}, C.N. Pope\hoch{\ddagger2},
J.F. V\'azquez-Poritz\hoch{\dagger1}}

\begin{document}
\begin{flushright}
\hfill{CTP TAMU-16/00}\\
\hfill{UPR-891-T}\\
\hfill{hep-th/0005246}\\
\hfill{May, 2000}\\
\end{flushright}


\begin{center}
{ \large {\bf AdS in Warped Spacetimes}}

\vspace{15pt}
\auth

\vspace{15pt}

{\hoch{\dagger}\upenn}

\vspace{15pt}
{\hoch{\ddagger}\tamphys}

\vspace{40pt}

\underline{ABSTRACT}
\end{center}

         We obtain a large class of AdS spacetimes warped with certain
internal spaces in eleven-dimensional and type IIA/IIB supergravities.
The warp factors depend only on the internal coordinates. These
solutions arise as the near-horizon geometries of more general
semi-localised multi-intersections of $p$-branes.  We achieve this by
noting that any sphere (or AdS spacetime) of dimension greater than 3
can be viewed as a foliation involving $S^3$ (or AdS$_3$).  Then the
$S^3$ (or AdS$_3$) can be replaced by a three-dimensional lens space
(or a BTZ black hole), which arises naturally from the introduction of
a NUT (or a pp-wave) to the M-branes or the D3-brane.  We then
obtain multi-intersections by performing a Kaluza-Klein reduction or
Hopf T-duality transformation on the fibre coordinate of the lens
space (or the BTZ black hole).  These geometries provide further
possible examples of the AdS/CFT correspondence and of consistent
embeddings of lower-dimensional gauged supergravities in $D=11$ or
$D=10$.

{\vfill\leftline{}\vfill
\footnoterule
{\footnotesize \hoch{1} Research supported in part by DOE grant 
DE-FG02-95ER40893 \vskip -12pt} \vskip 14pt
{\footnotesize  \hoch{2} Research supported in part by DOE 
grant DE-FG03-95ER40917.\vskip  -12pt}}

\pagebreak
\setcounter{page}{1}

\section{Introduction}

       Anti-de Sitter (AdS) spacetimes naturally arise as the
near-horizon geometries of non-dilatonic $p$-branes in supergravity
theories.  The metric for such a solution is usually the direct sum of
AdS and an internal sphere.  These geometries are of particular
interest because of the conjecture that supergravity on such a
background is dual to a conformal field theory on the boundary of the
AdS \cite{malda,gkp,wit}.  Examples include all the anti-de Sitter
spacetimes AdS$_d$ with $2\le d\le 7$, with the exception of $d=6$.
The origin of AdS$_6$ is a little more involved, and it was first
suggested in \cite{fkpz} that it was related to the ten-dimensional
massive type IIA theory.  Recently, it was shown that the massive type
IIA theory admits a warped-product solution of AdS$_6$ with $S^4$
\cite{bo}, which turns out to be the near-horizon geometry of a
semi-localised D4/D8 brane intersection \cite{youm}.  It is important
that the warp factors depend only on the internal $S^4$ coordinates,
since this implies that the reduced theory in $D=6$ has AdS spacetime
as its vacuum solution.  The consistent embedding of $D=6$, $N=1$
gauged supergravity in massive type IIA supergravity was obtained in
\cite{d6gauge}.  Ellipsoidal distributions of the D4/D8 system were
also obtained, giving rise to AdS domain walls in $D=6$, supported by
a scalar potential involving 3 scalars \cite{dist}.

   In fact, configurations with AdS in a warped spacetime are not rare
occurrences.  In \cite{oz}, a semi-localised M5/M5 system \cite{youm}
was studied, and it was shown that the near-horizon geometry turns out
to be a warped product of AdS$_5$ with an internal 6-space.  This
makes it possible to study AdS$_5$/CFT$_4$ from the point of view of
M-theory.  In this paper, we shall consider AdS with a warped
spacetime in a more general context and obtain such geometries for
all the AdS$_d$, as the near-horizon limits of semi-localised multiple
intersections in both type IIA and type IIB theories.

   The possibility of this construction is based on the following
observations.  As is well known, a non-dilatonic $p$-brane has the
near-horizon geometry AdS$_d \times S^n$. The internal $n$-sphere can
be described geometrically as a foliation of $S^p\times S^q$ surfaces
with $n=p+q+1$ (see appendix A), and so, in particular, if $n\ge 4$
the $n$-sphere can be viewed in terms of a foliation with $S^3\times
S^{n-4}$ surfaces, {\it viz.}
\be
d\Omega_n^2 = d\a^2 + \cos^2\a\, d\Omega_3^2 + \sin^2\a\, 
d\Omega_{n-4}^2\,.\label{folia}
\ee
In appendix B, we show that when a non-dilatonic $p$-brane with an
$n$-sphere in the transverse space intersects with a Kaluza-Klein
monopole (a Taub-NUT with charge $\Qn$) in a semi-localised manner,
the net result turns out to be effectively a coordinate transformation
of a solution with a distribution of pure $p$-branes with no NUT
present.  The round $S^3$ in (\ref{folia}) becomes the cyclic lens
space $S^3/Z_{\Qn}$ with metric
\be
d\bOmega_3^2 = \ft14 d\Omega_2^2 + \ft14 (\fft{dy}{\Qn} + \omega)^2\,,
\label{lens}
\ee 
where $d\omega=\Omega_2$ is the volume form of the unit 2-sphere.
This metric retains the same local structure as the standard round
3-sphere, and it has the same curvature tensor, but the $y$ coordinate
on the $U(1)$ fibres is now identified with a period which is $1/\Qn$
of the period for $S^3$ itself.  We can now perform a dimensional
reduction, or a T-duality transformation, on the fibre coordinate $y$,
and thereby obtain AdS in a warped spacetime.  The warp factor depends
only on the internal ``latitude'' coordinate $\a$, but is independent
of the lower-dimensional spacetime coordinates.  In fact, the M5/M5
system with AdS$_5$ found in \cite{oz} can be obtained in precisely
such a manner from the D3-brane by using type IIA/IIB T-duality.  Note
that an isotropic $p$-brane can be viewed as carrying a single unit of
NUT charge.  Although this semi-localised way of introducing a
Taub-NUT seems trivial, in that it amounts to a coordinate
transformation, performing Kaluza-Klein reduction on the fibre
coordinate does create a non-trivial intersecting component, since the
Kaluza-Klein 2-form field strength now carries a non-trivial
flux. This fact was used in \cite{create} to construct multi-charge
$p$-branes starting from flat spacetime.

    An analogous procedure can instead be applied to the anti-de
Sitter spacetime, rather than the sphere, in the near-horizon limit
AdS$_d\times S^n$ of a non-dilatonic $p$-brane.  As discussed in
appendix A, AdS$_d$ can be described in terms of a foliation of
AdS$_p\times S^q$ surfaces with $d=p+q+1$ and so, in particular, for
$d\ge 4$ it can be expressed as a foliation of AdS$_3\times S^{d-3}$:
\be 
ds_{\rm{AdS}_d}^2 = d\rho^2 + \cosh^2\rho\, ds_{\rm{AdS}_3}^2 +
\sinh^2\rho\, d\Omega_{d-4}^2\,.  
\ee
In the presence of a pp-wave that is semi-localised on the
world-volume of the $p$-brane, the AdS$_3$ turns out to have the
form of a $U(1)$ bundle over AdS$_2$ \cite{s3twist}, 
\be 
ds_{\rm{AdS_3}}^2 = -r^2\,W^{-1}\, dt^2 + \fft{dr^2}{r^2} + r^2\,
W\, (dy + (W^{-1}-1)) dt)^2\,,\label{ads3a}
\ee
where $W=1 + Q_w/r^2$, and $Q_w$ is the momentum carried by the
pp-wave.  This is precisely the structure of the extremal BTZ black
hole \cite{btz}.  We can now perform a Kaluza-Klein reduction, or
T-duality transformation, on the fibre coordinate $y$.  In the
near-horizon limit where the ``1'' in $W$ can be dropped, we obtain
AdS$_2$ in a warped spacetime with a warp factor that depends only on
the foliation coordinate, $\rho$.

    A T-duality transformation on such a fibre coordinate of AdS$_3$
or $S^3$ has been called Hopf T-duality \cite{s5twist}.  It has the
effect of (un)twisting the AdS$_3$ or $S^3$.  The effect of this
procedure on the six-dimensional dyonic string, whose near-horizon
limit is AdS$_3\times S^3$, was extensively studied in \cite{s3twist}.
In this paper, we apply the same technique to AdS$_3$ or $S^3$
geometries that are themselves factors in the foliation surfaces of
certain larger-dimensional AdS spacetimes or spheres.  

        In section 2, we consider the semi-localised D3/NUT system and
show that the effect of turning on the NUT charge $\Qn$ in the
intersection is merely to convert the internal 5-sphere, viewed
as a foliation of $S^1\times S^3$, into a foliation of $S^1\times
(S^3/Z_{\Qn})$, where $S^3/Z_{\Qn}$ is the cyclic lens space of order
$\Qn$.  We can then perform a T-duality transformation on the Hopf
fibre coordinate of the lens space and thereby obtain an AdS$_5$ in a
warped spacetime as a solution in M theory, as the near-horizon
geometry of a semi-localised M5/M5 system.

        In section 3, we consider a semi-localised D3/pp-wave system,
for which the AdS$_5$ becomes a foliation of a circle with the
extremal BTZ black hole, which is locally AdS$_3$ and can be viewed as
a $U(1)$ bundle over AdS$_2$.  We then perform a Hopf T-duality
transformation on the fibre coordinate to obtain a solution with
AdS$_2$ in a warped spacetime in M-theory, as the near-horizon
geometry of a semi-localised M2/M2 system.

         In sections 4 and 5, we apply the same analysis to the M2/NUT
and M2/pp-wave systems, and the M5/NUT and M5/pp-wave systems,
respectively; we obtain various configurations of AdS in warped
spacetimes by performing Kaluza-Klein reductions and Hopf T-duality
transformations on the fibre coordinates.

         In section 6, we consider the D4/D8 system, which has the
near-horizon geometry of a warped product of AdS$_6$ and $S^4$.  We
perform a Hopf T-duality transformation on the fibre coordinate of the
foliating lens space of $S^4$, and thereby embed AdS$_6$ in a
warped spacetime solution of type IIB theory.  

          We end with concluding remarks in section 7.  In appendix A,
we show how arbitrary-dimensional spheres and AdS spacetimes can be
described in terms of foliations.  In appendix B, we show that the
solution describing the semi-local intersection of a non-dilatonic
$p$-brane with a Kaluza-Klein monopole (Taub-NUT) is equivalent, after
a coordinate transformation, to a solution purely composed of
distributed $p$-branes, with no NUT.

\section{D3/NUT systems and AdS$_5$ in M-theory from T-duality}

      AdS$_5$ spacetime arises naturally from type IIB theory as the
near-horizon geometry of the D3-brane.  Its origin in M-theory is more
obscure.  One way to embed the AdS$_5$ in M-theory is to note that
$S^5$ can be viewed as a $U(1)$ bundle over $CP^2$, and hence we can
perform a Hopf T-duality transformation on the $U(1)$ fibre
coordinate.  The resulting M-theory solution becomes AdS$_5\times
CP^2\times T^2$ \cite{s5twist}.  However, this solution is not
supersymmetric at the level of supergravity, since $CP^2$ does not
admit a spin structure.  {\it Charged} spinors exist but, after making
the T-duality transformation, the relevant electromagnetic field is
described by the winding-mode vector and it is only in the full
string theory that states charged with respect to this field arise. It
was therefore argued in \cite{s5twist} that the lack of supersymmetry
(and indeed of any fermions at all) is a supergravity artifact and
that, when the full string theory is considered, the geometry is
supersymmetric.  Such a phenomenon was referred as ``supersymmetry
without supersymmetry'' in \cite{s7twist}.

        Recently, AdS$_5$ in warped eleven-dimensional spacetime was
constructed in \cite{oz}.  It arises as the near-horizon limit of the
semi-localised M5/M5 intersecting system. After performing a T-duality
transformation, the warped spacetime of the near-horizon limit becomes
AdS$_5\times (S^5/Z_{\Qn})$. In this section, we shall review this
example in detail and show that the M5/M5 system originates from a
semi-localised D3/NUT intersection in type IIB supergravity.

\subsection{D3/NUT system}

        Any $p$-brane with a transverse space of sufficiently high
dimension can intersect with a NUT.  The D3/NUT solution of type IIB
supergravity is given by
\bea
ds_{10\rm{IIB}}^2 &=& H^{-1/2}(-dt^2 + dw_1^2 + \cdots + dw_3^2) +
H^{1/2}\Big(dx_1^2 + dx_2^2\nn\\
&&\phantom{xxxxxx} K(dz^2+ z^2\, d\Omega_2^2) + K^{-1}(dy +
\Qn\, \omega)^2 \Big)\,,\label{d3nut}\\
F_5&=&dt\wedge d^3w\wedge dH^{-1} + {*(dt\wedge d^3w\wedge dH^{-1})}
\,,\nn
\eea
where $z^2 = z_1^2 + z_2^2 + z_3^2$, and $\omega$ is a 1-form
satisfying $d\omega = \Omega_2$.  The solution can be best
illustrated by the following diagram:

\bigskip\bigskip
\centerline{
\begin{tabular}{c|cccccccccccc}
&$t$ & $w_1$ & $w_2$ & $w_3$ & $x_1$ & $x_2$ & $z_1$ & $z_2$ &
$z_3$ & $y$ & \\ \hline
D3&$\times$ & $\times$ & $\times$ & $\times$ &$-$ &$-$ &$-$ &$-$ &$-$
&$-$ & $H$ \\
NUT&$\times$ & $\times$ & $\times$ & $\times$ & $\times$ & $\times$
&$-$ &$-$ &$-$ &$*$ & $K$ \\
\end{tabular}}
\bigskip

\centerline{Diagram 1.  The D3/NUT brane intersection.  Here
$\times$ and $-$ denote the}
\centerline{$\phantom{Diagram 1. }$
worldvolume and transverse space coordinates respectively,}
\centerline{$\phantom{xxxx}$ and $*$ denotes the fibre coordinate 
of the Taub-NUT.}
\bigskip\bigskip

        The function $K$ is associated with the NUT component of the
intersection; it is a harmonic function in the overall transverse
Euclidean 3-space coordinatised by $z_i$.  The function $H$ is
associated with the D3-brane component.  It satisfies the equation
\be
\del_{\vec z}^2 H + K\, \del_{\vec x}^2 H=0\,.
\ee
Equations of this type were also studied in
\cite{cvet1,cvet2,tseyt1,horo,callan,tseyt2,lpint,ity,02128,02210,03038}.
In the absence of NUT charge, {\it i.e.} $K=1$, the function $H$ is
harmonic in the the transverse 6-space of the D3-brane.  When the NUT
charge $\Qn$ is non-zero, $K$ is instead given by
\be
K=1 + \fft{\Qn}{z}\,,\label{knut}
\ee
and the function $H$ cannot be solved analytically, but only in terms
of a Fourier expansion in $\vec x$ coordinates.  The usual way to
solve for the solution is to consider the zero-modes in the Fourier
expansion.  In other words, one assumes that $H$ is independent of
$\vec x$.  The consequence of this assumption is that the resulting
metric no longer has an AdS structure in its near-horizon region.  In
\cite{youm}, it was observed that an explicit closed-form solution for
$H$ can be obtained in the case where the ``1'' in function K is
dropped.  This solution is given by \cite{youm}
\be 
K = \fft{\Qn}{z}\,,\qquad H = 1 + \sum_k\fft{Q_k}{(|\vec x - \vec
x_{0k}|^2 + 4\Qn\, z)^2} \,.
\ee
In this paper, we shall consider the case where the D3-brane is
located at the origin of the $\vec x$ space and so we have
\be
H = 1 + \fft{Q}{(x^2 + 4 \Qn\, z)^2}\,,
\ee
where $x^2 = x^i \,x^i$.  Thus, the D3-brane is also localised in the
space of the $\vec x$ as well.  Let us now make a coordinate
transformation
\be
x_1=r\, \cos\a\, \cos\theta\,,\quad
x_2=r\, \cos\a\, \sin\theta\,,
\quad z = \ft14 \Qn^{-1}\, r^2\, \sin^2\a\,.
\label{d2nutvar}
\ee
In terms of the new coordinates, the metric for the solution becomes
\bea
ds_{10\rm{IIB}}^2 &=& H^{-1/2}(-dt^2 + dw_1^2 + dw_2^2 + dw_3^2) +
H^{1/2}(dr^2 + r^2\, dM_5^2)\,,\nn\\
H&=&1 + \fft{Q}{r^4}\,.
\eea
where
\be
dM_5^2 = d\a^2 + c^2\,d\theta^2 + \ft14s^2\Big(d\Omega_2^2
+(\fft{dy}{\Qn} + \omega)^2\Big)\,,
\ee
and $s=\sin\a$, $c=\cos\a$.  Thus, we see that $dM_5^2$ describes a
foliation of $S^1$ times the lens space $S^3/Z_{\Qn}$.  For a unit NUT
charge, $\Qn=1$, the metric $dM_5^2$ describes the round 5-sphere and
the solution becomes an isotropic D3-brane.  It is interesting to note
that the regular D3-brane can be viewed as a semi-localised D3-brane
intersecting with a NUT with unit charge.\footnote{An analogous
observation was also made in \cite{create}, where multi-charge
solutions were obtained from flat space by making use of the fact that
$S^3$ can be viewed as a $U(1)$ bundle over $S^2$.  In other words,
flat space can be viewed as a NUT, with unit charge, located on the
$U(1)$ coordinate.}  In the near-horizon limit $r\rightarrow 0$, where
the constant 1 in the function $H$ can be dropped, the metric becomes
AdS$_5\times M_5$:
\be
ds_{10\rm{IIB}}^2 = Q^{-1/2}\, r^2\, (-dt^2 + dw^i dw^i) +
Q^{1/2}\fft{dr^2}{r^2} + Q^{1/2}\Big(d\a^2 + c^2\,d\theta^2 + 
\ft14s^2(d\Omega_2^2
+(\fft{dy}{\Qn} + \omega)^2)\Big)\,.\label{d3nuthorizon}
\ee

\subsection{M5/M5 system and AdS$_5$ in M-theory}

      Since the near-horizon limit of a semi-localised D3-brane/NUT is
a direct product of AdS$_5$ and an internal 5-sphere that is a
foliation of a circle times a lens space, it follows that if we
perform a T-duality transformation on the $U(1)$ fibre coordinate $y$,
we shall obtain AdS$_5$ in a warped spacetime as a solution of the
type IIA theory.  The warp factor is associated with the scale factor
$s^2$ of $dy^2$ in (\ref{d3nuthorizon}). This type of Hopf T-duality
has the effect of untwisting a 3-sphere into $S^2\times S^1$
\cite{s3twist}.  If one performs the T-duality transformation on the
original full solution (\ref{d3nut}), rather than concentrating on its
near-horizon limit, then one obtains a semi-localised NS5/D4 system of
the type IIA theory, which can be further lifted back to $D=11$ to
become a semi-localised M5/M5 system, obtained in \cite{youm}.  In
\cite{oz}, the near-horizon structures of these semi-localised branes
of M-theory were analysed, and AdS$_5$ was obtained as a warped
spacetime solution.  We refer the readers to Ref.~\cite{oz} and shall
not discuss this solution further, but only mention that, from the
above analysis, it can be obtained by implementing the T-duality
transformation on the coordinate $y$ in (\ref{d3nuthorizon}).

\section{D3/pp-wave system and extremal BTZ black hole}

       In this section, we study the semi-localised pp-wave
intersecting with a D3-brane.  The solution is given by
\bea
ds_{10\rm{IIB}}^2 &=& H^{-1/2}\Big(-W^{-1}\, dt^2 + 
W\,(dy + (W^{-1}-1)dt)^2 +dx_1^2 + dx_2^2\Big)\nn\\
&&+ H^{1/2} (dz_1^2 + \cdots dz_6^2)\,,\label{d3wave}\\
F_\5&=&dt\wedge dy\wedge dx_1\wedge dx_2\wedge dH^{-1} +
{*(dt\wedge dy\wedge dx_1\wedge dx_2\wedge dH^{-1})}\,,\nn
\eea
The solution can be illustrated by the following diagram

\bigskip\bigskip
\centerline{
\begin{tabular}{c|cccccccccccc}
&$t$ & $y$ & $x_1$ & $x_2$ & $z_1$ & $z_2$ & $z_3$ & $z_4$ &
$z_5$ & $z_6$ & \\ \hline
D3&$\times$ & $\times$ & $\times$ & $\times$ &$-$ &$-$ &$-$ &$-$ &$-$
&$-$ & $H$ \\
wave&$\times$ & $\sim$ & $-$ & $-$ & $-$ & $-$
&$-$ &$-$ &$-$ &$-$ & $W$ \\
\end{tabular}}
\bigskip

\centerline{Diagram 2.  The D3/pp-wave brane intersection.  Here
$\sim$ denotes the wave coordinate.}

\bigskip\bigskip

    In the usual construction of such an intersection, the harmonic
functions $H$ and $W$ depend only on the overall transverse space
coordinates $\vec z$.  The near-horizon limit of the solution then
becomes K$_5\times S^6$, where K$_5$ is the generalised Kaigorodov
metric in $D=5$, and the geometry is dual to a conformal field theory
in the infinite momentum frame \cite{kaig}.  On the other hand, the
semi-localised solution is given by \cite{youm}
\be
H=\fft{Q}{|\vec z|^4}\,,\qquad
W= 1 + Q_w (|\vec x|^2 + \fft{Q}{|\vec z|^2})\,.
\ee
We now let
\be
x_1 = \fft{1}{r}\,\cos\a\, \cos\theta\,,\quad
x_2 = \fft{1}{r}\,\cos\a\, \sin\theta\,,\qquad
z_i = \fft{r\, Q^{1/2}}{\sin\a}\, \nu_i\,,\label{d3wavenewcord}
\ee
where $\nu_i$ coordinates, satisfying $\nu_i\,\nu_i=1$, define a
5-sphere with the unit sphere metric $d\Omega_5^2 = d\nu_i\, d\nu_i$.
Using these coordinates, the metric of the semi-localised D3/wave
system becomes
\be
ds_{10\rm{IIB}}^2 = Q^{1/2}\, s^{-2}\, (ds_{\rm{AdS}_3}^2 + d\a^2 + 
c^2\, d\theta^2 +s^2\, d\Omega_5^2)\,,\label{d3wavehorizon}
\ee
where $ds_{\rm{AdS}_3}^2$ is given by
\bea
ds_{\rm{AdS}_3}^2 &=& -r^2\, W^{-1}\, dt^2 + r^2\,W\, (dy + (W^{-1}
-1)dt)^2 + \fft{dr^2}{r^2}\,,\nn\\
W&=& 1+ \fft{Q_w}{r^2}\,.\label{ads3}
\eea
Note that the above metric is exactly the extremal BTZ black hole
\cite{btz}, and hence it is locally AdS$_3$.  Thus we have
demonstrated that the semi-localised D3/pp-wave system is in fact a
warped product of AdS$_3$ (the extremal BTZ black hole) with a
7-sphere, where $S^7$ is described as a foliation of $S^1\times S^5$
surfaces.\footnote{A D3-brane with an $S^3\times \R$ worlvolume was
obtained in \cite{papadop}.  In that solution, which was rather
different from ours, the dilaton was not constant.}  Note that the
metric (\ref{d3wavehorizon}) can also be expressed as a direct product
of AdS$_5\times S^5$, with the AdS$_5$ metric written in the following
form:
\be
ds_5^2 = s^{-2}(ds_{\rm{AdS}_3}^2 + d\a^2 + c^2\, d\theta^2)\,.
\label{ads5a}
\ee
Making a coordinate transformation $\tan(\a/2) = e^\rho$, the metric
becomes
\be
ds_5^2 = d\rho^2 + \sinh^2\rho\, d\theta^2 +
\cosh^2\rho\, ds_{\rm{AdS}_3}^2\,,
\ee
which is precisely the AdS$_5$ metric written as a foliation of a
circle times AdS$_3$ (see appendix A).

    The extremal BTZ black hole occurs \cite{ss} as the near-horizon
geometry of the boosted dyonic string in six-dimensions, which can be
viewed as an intersection of a string and a 5-brane in $D=10$.  The
boosted D1/D5 system was used to obtain the first stringy
interpretation \cite{sv} of the microscopic entropy of the
Reissner-Nordstr\"om black hole in $D=5$.  The boosted dyonic string
has three parameters, namely the electric and magnetic charges $Q_e$,
$Q_m$, and the boost momentum parameter $Q_w$.  On the other hand, the
extremal BTZ black hole itself has only two parameters: the
cosmological constant, proportional to $\sqrt{Q_e\, Q_m}$, and the
mass (which is equal to the angular momentum in the extremal limit),
which is related to $Q_w$.  (Analogous discussion applies to $D=4$
\cite{cl}.)  In our construction of the BTZ black hole in warped
spacetime, the original configuration also has only two parameters,
namely the D3-brane charge $Q$, related to the cosmological constant
of the BTZ black hole, and the pp-wave charge, associated with the
mass.

\subsection{NS1/D2 and M2/M2 systems and AdS$_2$}

          We can perform a T-duality transformation on the coordinate
$y$ in the previous solution.  The D3-brane is T-dual to the D2-brane,
and the wave is T-dual to the NS-NS string.  Thus the D3/pp-wave
system of the type IIB theory becomes an NS1/D2 system in the type IIA
theory, given by
\bea
ds_{10\rm{IIA}}^2 &=& W^{1/4}H^{3/8} \Big[-(WH)^{-1}\, dt^2 +
H^{-1}\,(dx_1^2+ dx_2^2) + W^{-1}\,dy_1^2\,,\nn\\
&&\phantom{xxxxxxxxxxx} + dz_1^2 + \cdots dz_6^2\Big]\,,\nn\\
e^{\phi} &=& W^{-1/2}\, H^{1/4}\,,\label{ns1d2}\\
F_\4 &=& dt\wedge dx_1\wedge dx_2\wedge dH^{-1}\,,\qquad
F_\3 = dt\wedge dy_1\wedge dW^{-1}\,.\nn
\eea
This solution can be represented diagrammatically as follows:

\bigskip\bigskip
\centerline{
\begin{tabular}{c|ccccccccccc}
&$t$ & $x_1$ & $x_2$ & $y_1$ & $z_1$ & $z_2$ & $z_3$ & $z_4$ &
$z_5$ & $z_6$ & \\ \hline
D2&$\times$ & $\times$ & $\times$ & $-$ &$-$ &$-$ &$-$ &$-$ &$-$
&$-$ & $H$ \\
NS1&$\times$ & $-$ & $-$ & $\times$ & $-$ &$-$ &$-$ &$-$ &$-$
&$-$ & $W$ \\
\end{tabular}}
\bigskip

\centerline{Diagram 3.  The NS1/D2 brane intersection.}
\bigskip\bigskip

          In the near-horizon limit where the 1 in $W$ is
dropped, the metric of the NS1/D2 system (\ref{ns1d2}), in terms of
the new coordinates (\ref{d3wavenewcord}), becomes
\be
ds_{10}^2 =Q_w^{1/4}Q^{5/8}\,s^{-5/2}\, \Big(ds_{\rm{AdS}_2}^2 + 
d\a^2 + c^2\, d\theta^2 + s^2\, d\Omega_5^2
+ (Q_w\,Q)^{-1}\,s^4\, dy_1^2\,\Big)\,,
\ee
where
\be
ds_{\rm{AdS}_2}^2 =-\fft{r^4\, dt^2}{Q_w} +\fft{dr^2}{r^2}\,.
\label{ads2}
\ee
Thus we see that the near-horizon limit of the NS1/D2 system is a
warped product of AdS$_2$ with a certain internal 8-space, which is a
warped product of a 7-sphere with a circle.

            We can further lift the solution back to $D=11$, where it
becomes a semi-localised M2/M2 system,
\bea
ds_{11}^2 &=& (WH)^{1/3}\Big[-(WH)^{-1}\, dt^2 +
H^{-1}\,(dx_1^2 + dx_2^2) + W^{-1}\,(dy_1^2 + dy_2^2)\,,\nn\\
&&\phantom{xxxxxxxxxxx} + dz_1^2 + \cdots +dz_6^2\Big]\,,\nn\\
F_\4 &=& dt\wedge dx_1\wedge dx_2\wedge dH^{-1} +
       dt\wedge dy_1\wedge dy_2\wedge dW^{-1}\,.\label{m2m2}
\eea
The configuration for this solution can be summarised in the following
diagram: 

\bigskip\bigskip
\centerline{
\begin{tabular}{c|cccccccccccc}
&$t$ & $x_1$ & $x_2$ & $y_1$ & $y_2$ & $z_1$ & $z_2$ & $z_3$ & $z_4$ &
$z_5$ & $z_6$ & \\ \hline
M2&$\times$ & $\times$ & $\times$ & $-$ &$-$ &$-$ &$-$ &$-$ &$-$ &$-$
&$-$ & $H$ \\
M2&$\times$ & $-$ & $-$ & $\times$ & $\times$ & $-$ &$-$ &$-$ &$-$ &$-$
&$-$ & $W$ \\
\end{tabular}}
\bigskip

\centerline{Diagram 4.  The M2-M2 brane intersection.}
\bigskip\bigskip

   It is straightforward to verify that the near-horizon geometry of
this system is a warped product of AdS$_2$ with a certain 9-space,
namely
\be
ds_{11}^2 = Q_w^{1/3}Q^{2/3}\,s^{-8/3}\, 
(ds_{\rm{AdS}_2}^2 + d\a^2 + c^2\, d\theta^2 + s^2\, d\Omega_5^2
+ (Q_w\,Q)^{-1}\,s^4\, (dy_1^2 + dy_2^2))\,,
\ee
where $ds_{\rm{AdS}_2}^2$ is an AdS$_2$ metric given by (\ref{ads2}),
and the internal 9-space is a warped product of a 7-sphere and a
2-torus.

\subsection{Further possibilities}

       Note that in the above examples, we can replace the round
sphere $d\Omega_5^2$ by a lens space of the following form:
\be
d\Omega_5^2 = d\td \a^2 + \td c^2\, d\td \theta^2 + \td s^2
\, (d\wtd \Omega_2^2 + (\fft{d\td y}{\wtd Q_{\sst{\rm N}}} + 
\td \omega)^2)\,,
\ee
where $\td c\equiv \cos\td \a$, $\td s\equiv \sin\td \a$ and $d\td
\omega = \wtd \Omega_2$.  As we have discussed in appendix B, this can
be viewed as an additional NUT with charge $\wtd Q_{\sst{\rm N}}$
intersecting with the system.  We can now perform a Kaluza-Klein
reduction or T-duality transformation on the fibre coordinate $\td y$,
leading to many further examples of warped products of AdS$_2$ or
AdS$_3$ with certain internal spaces.  The warp factors again depend
only on the coordinates of the internal space.  These geometries can
be viewed as the near-horizon limits of three intersecting branes,
with charges $Q$, $\Qn$ and $\wtd Q_{\sst{\rm N}}$.  Of course, this
system can equally well be obtained by replacing the horospherical
AdS$_5$ in (\ref{d3nuthorizon}) with (\ref{ads5a}).

       For example, let us consider the M2/M2 system with an
additional NUT component.  The solution of this semi-localised
intersecting system is given by
\bea
ds_{11}^2 &=& (WH)^{1/3}\Big[-(WH)^{-1}\, dt^2 +
H^{-1}\,(dx_1^2 + dx_2^2) + W^{-1}\,(dy_1^2 + dy_2^2)\,,\nn\\
&&\phantom{xxxxxxxxxxx} + K(dz^2 + z^2\, d\Omega_2^2) + 
K^{-1}(dy + \Qn\,\omega)^2 + du_1^2 + du_2^2\Big]\,,\nn\\
F_\4 &=& dt\wedge dx_1\wedge dx_2\wedge dH^{-1} +
       dt\wedge dy_1\wedge dy_2\wedge dW^{-1}\,.\label{m2m2nut}
\eea
where the functions $H$, $W$ and $K$ are given by
\be
H=\fft{Q}{(|\vec u|^2 + 4\Qn\, z)^2}\,,\quad
W=1 + Q_w(|\vec x|^2 + \fft{Q}{|\vec u|^2 + 4\Qn\, z})\,,\quad
K=\fft{\Qn}{z}\,.
\ee
We illustrate this solution in the following diagram:

\bigskip\bigskip
\centerline{
\begin{tabular}{c|cccccccccccc}
&$t$ & $x_1$ & $x_2$ & $y_1$ & $y_2$ & $z_1$ & $z_2$ & $z_3$ & $y$ &
$u_1$ & $u_2$ & \\ \hline
M2&$\times$ & $\times$ & $\times$ & $-$ &$-$ &$-$ &$-$ &$-$ &$-$ &$-$
&$-$ & $H$ \\
M2&$\times$ & $-$ & $-$ & $\times$ & $\times$ & $-$ &$-$ &$-$ &$-$ &$-$
&$-$ & $W$ \\
NUT& $\times$ & $\times$ & $\times$ & $\times$ & $\times$ & $-$
& $-$ & $-$ & $*$ &$\times$ & $\times$ & $K$
\end{tabular}}
\bigskip

\centerline{Diagram 5.  The M2/M2/NUT brane intersection.}
\bigskip\bigskip

          The near-horizon structure of this solution is basically the
same as that of the M2/M2 system with the round $S^3$ in the foliation
replaced by the lens space $S^3/Z_{\Qn}$.  We can now perform
Kaluza-Klein reduction on the fibre coordinate $y$ and the solution
becomes the semi-localised D2/D2/D6 brane intersection, given by
\bea
ds_{10\rm{IIA}}^2 &=& (WH)^{3/8}K^{-1/8}\,\Big[-(WH)^{-1}\, dt^2 +
H^{-1}\,(dx_1^2 + dx_2^2) + W^{-1}\,(dy_1^2 + dy_2^2)\,,\nn\\
&&\phantom{xxxxxxxxxxx} + K(dz^2 + z^2\, d\Omega_2^2) + 
du_1^2 + du_2^2\Big]\,,\nn\\
F_\4 &=& dt\wedge dx_1\wedge dx_2\wedge dH^{-1} +
       dt\wedge dy_1\wedge dy_2\wedge dW^{-1}\,.\label{d2d2d6}\\
e^{\phi}&=& (W\, H)^{1/4} K^{-3/4}\,,\qquad
F_\2 = \Qn\,\Omega_2\,.
\eea
The solution can be illustrated by the following diagram:

\newpage
\centerline{
\begin{tabular}{c|ccccccccccc}
&$t$ & $x_1$ & $x_2$ & $y_1$ & $y_2$ & $z_1$ & $z_2$ & $z_3$ &
$u_1$ & $u_2$ & \\ \hline
D2&$\times$ & $\times$ & $\times$ & $-$ &$-$ &$-$ &$-$ &$-$ &$-$
&$-$ & $H$ \\
D2&$\times$ & $-$ & $-$ & $\times$ & $\times$ & $-$ &$-$ &$-$ &$-$
&$-$ & $W$ \\
D6& $\times$ & $\times$ & $\times$ & $\times$ & $\times$ & $-$
& $-$ & $-$ &$\times$ & $\times$ & $K$
\end{tabular}}
\bigskip

\centerline{Diagram 6.  The D2/D2/D6 brane intersection.}
\bigskip\bigskip

\section{M2/NUT and M2/pp-wave systems}

       In this section, we apply an analogous analysis to the
M2-brane.  We show that the semi-localised M2-brane intersecting with
a NUT is in fact an isotropic M2-brane with the internal 7-sphere
itself being described as a foliation of a regular $S^3$ and lens
space $S^3/Z_{\Qn}$, where $\Qn$ is the NUT charge.  Reducing the
system to $D=10$, we obtain a semi-localised D2/D6 system whose
near-horizon geometry is a warped product of AdS$_4$ with an internal
6-space.  We also show that a semi-localised pp-wave intersecting with
the M2-brane is in fact a warped product of AdS$_3$ (the BTZ black
hole) and an 8-space.  The system can be reduced to $D=10$ to become a
semi-localised D0/NS1 intersection.

\subsection{M2-brane/NUT system}

         The solution for the intersection of an M2-brane and a NUT is
given by
\bea
ds_{11}^2 &=& H^{-2/3}\,(-dt^2 + dw_1^2 + dw_2^2) + H^{1/3}\Big(dx_1^2 +
\cdots + dx_4^2\nn\\
&&\phantom{xxxx}+ K(dz^2 + z^2 d\Omega_2^2) + K^{-1} (dy + \Qn\, \omega)^2
\Big)\,,\nn\\
F_\4 &=& dt\wedge dw_1\wedge dw_2\wedge dH^{-1}\,,\label{m2nut}
\eea
where $z^2 = z_1^2 + z_2^2 + z_3^2$ and $d\omega = \Omega_2$.  
The solution can be illustrated by the following diagram:

\bigskip\bigskip
\centerline{
\begin{tabular}{c|cccccccccccc}
&$t$ & $w_1$ & $w_2$ & $x_1$ & $x_2$ & $x_3$ & $x_4$ & $z_1$ & $z_2$ &
$z_3$ & $y$ & \\ \hline
M2&$\times$ & $\times$ & $\times$ & $-$ &$-$ &$-$ &$-$ &$-$ &$-$ &$-$
&$-$ & $H$ \\
NUT&$\times$ & $\times$ & $\times$ & $\times$ & $\times$ & $\times$
&$\times$ &$-$ &$-$ &$-$ &$*$ & $K$ \\
\end{tabular}}
\bigskip

\centerline{Diagram 7.  The M2/NUT brane intersection.}
\bigskip\bigskip
 
        If the function $K$ associated with the NUT components of the
intersection takes the form $K=\Qn/z$, then the function $H$
associated with the M2-brane component can be solved in the
semi-localised form
\be
H = 1 + \fft{Q}{(|\vec x|^2 + 4 \Qn\, z)^3}\,.
\ee
Thus, the solution is also localised on the space of the $\vec x$
coordinates.  Let us now make a coordinate transformation
\be x_i=r\, \cos\a\, \mu_i,,\qquad z = \ft14 \Qn^{-1}\,
r^2\, \sin^2\a\,,
\label{m2nutvar}
\ee
where $\mu_i\,\mu_i =1$, defining a 3-sphere, with the unit 3-sphere
metric given by $d\Omega_3^2 = d\mu_i\, d\mu_i$.  In terms of the new
coordinates, the metric for the solution becomes
\bea
ds_{11}^2 &=& H^{-2/3} (-dt^2 + dw_1^2 + dw_2^2) + H^{1/3}(dr^2 + 
r^2\, dM_7^2)\,,\nn\\
H&=& 1 + \fft{Q}{r^6}\,,
\eea
where
\be
dM_7^2 = d\a^2 + c^2\, d\Omega_3^2 + \ft14 s^2\, 
\Big(d\Omega_2^2 + (\fft{dy}{\Qn} + \omega)^2\Big)\,.
\ee
Thus we see that $dM_7^2$ is a foliation of a regular 3-sphere,
together with a lens space $S^3/Z_{\Qn}$.  When $\Qn=1$ the metric
$dM_7^2$ describes a round 7-sphere and the solution becomes an
isotropic M2-brane.  Interestingly, the regular M2-brane can be viewed
as an intersecting semi-localised M2-brane with a NUT of unit
charge. In the near-horizon limit $r\rightarrow 0$, where the 1 in the
function $H$ can be dropped, the metric becomes AdS$_4\times M_7$.

\subsection{D2-D6 system}

      In the M2-brane and NUT intersection (\ref{m2nut}), we can
perform a Kaluza-Klein reduction on the $y$ coordinate.  This gives
rise to a semi-localised intersection of D2-branes and D6-branes:
\bea
ds_{10\rm{IIA}}^2 &=& H^{-5/8} K^{-1/8}\, (-dt^2 + dw_1^2 + dw_2^2) +
H^{3/8} K^{-1/8}\, (dx_1^2 + \cdots + dx_4^2)\nn\\
&&\phantom{xxxxxx} H^{3/8} K^{7/8}\, (dz_1^2 + dz_2^2 + dz_3^2)
\,,\nn\\
e^{\phi} &=& H^{1/4} K^{-3/4}\,,\label{d2d6}\\
F_\4&=&dt\wedge d^2w\wedge dH^{-1}\,,\qquad
F_2=e^{-3/2\phi} {*(dt\wedge d^2w\wedge d^4x\wedge dK^{-1})}\,.\nn
\eea
The solution can be illustrated by the following diagram

\newpage
\centerline{
\begin{tabular}{c|ccccccccccc}
&$t$ & $w_1$ & $w_2$ & $x_1$ & $x_2$ & $x_3$ & $x_4$ & $z_1$ & $z_2$ &
$z_3$ &\\ \hline
D2&$\times$ & $\times$ & $\times$ & $-$ &$-$ &$-$ &$-$ &$-$ &$-$
&$-$ & $H$ \\
D6&$\times$ & $\times$ & $\times$ & $\times$ & $\times$ & $\times$
&$\times$ &$-$ &$-$ &$-$ & $K$ \\
\end{tabular}}
\bigskip

\centerline{Diagram 8. The D2/D6 brane intersection.}
\bigskip\bigskip

   Again, in the usual construction of a D2-D6 system, the harmonic
functions $H$ and $K$ are taken to depend only on the overall
transverse space coordinates $\vec z$. In the semi-localized
construction, the function $H$ depends on $\vec x$ as well.  In terms
of the new coordinates defined in (\ref{m2nutvar}), the metric becomes
\be
ds_{10\rm{IIA}}^2 = (\fft{r\,s}{2\Qn})^{1/4}\Big[H^{-5/8}(-dt^2 + dw_1^2 +
dw_2^2) + H^{3/8}(dr^2 + r^2(d\a^2 + c^2\, d\Omega_3^2 + \ft14 s^2\,
d\Omega_2^2)\Big]\,.
\ee
Thus, in the near-horizon limit where the 1 in $H$ can be dropped, the
solution becomes a warped product of AdS$_4$ with an internal 6-space:
\be
ds_{10\rm{IIA}}^2 =(2\Qn)^{-1/4}Q^{3/8}\,s^{1/4}\,
(ds_{\rm{AdS}_4}^2 + d\a^2 + c^2\, d\Omega_3^2 + 
\ft14 s^2\, d\Omega_2^2)\,,
\ee
where $ds_4^2$ is the metric on AdS$_4$, given by
\be
ds_{\rm{AdS}_4}^2 = \fft{r^4}{Q}(-dt^2 + dw_1^2 +
dw_2^2) + \fft{dr^2}{r^2}\,.
\ee
The internal 6-space is a warped product of a 4-sphere with a 2-sphere.

\subsection{AdS$_4$ in type IIB from T-duality}

         In the above discussion, we found that our starting point is
effectively to replace the round 7-sphere of the M2-brane by the
foliation of a round 3-sphere together with a lens space
$S^3/Z_{\Qn}$.  We can also replace the round 3-sphere by another lens
space $S^3/Z_{\wtd Q_{\sst{\rm N}}}$, given by
\be
d\bOmega_3^2 = \ft14 \Big(d\wtd\Omega_2^2 + 
(\fft{d\td y}{\wtd Q_{\sst{\rm N}}} + \omega)^2\Big)\,.
\ee
As discussed in the appendix, the lens space arises from introducing a
NUT around the fibre coordinate $\td y$, with NUT charge $\wtd
Q_{\sst{\rm N}}$.  The system can then be viewed as the near-horizon
limit of three intersecting branes, with charges $Q$, $\Qn$ and $\wtd
Q_{\sst{\rm N}}$.  For example, with this replacement the D2/D6 system
becomes a D2/D6/NUT system. Performing a T-duality transformation on
the fibre coordinate $\td y$, the $S^3$ untwists to become $S^2\times
S^1$.  The resulting type IIB metric is given by
\be
ds_{10\rm{IIB}}^2 = 
\Big(\fft{Q\, s\, c}{4\Qn\,\wtd Q_{\sst{\rm N}}}\Big)^{1/2}\,
\Big(ds_{\rm{AdS}_4}^2 + d\a^2 + \ft14 c^2\, d\wtd\Omega_2^2 + 
\ft14 s^2\,d\Omega_2^2 + 
\fft{(4\Qn\,\wtd Q_{\sst{\rm N}})^2}{Q\, s^2\, c^2}\, d\td
y^2\Big)\,.\label{d3d5ns5horizon}
\ee
This metric can be viewed as describing the near-horizon geometry of
a semi-localised D3/D5/NS5 system in the type IIB theory.  This metric
(\ref{d3d5ns5horizon}) provides a background for consistent reduction
of type IIB supergravity to give rise to four-dimensional gauged
supergravity with AdS background.

         In order to construct the semi-localised D3/D5/NS5
intersecting system in the type IIB theory, we start with the
D2/D6/NUT system, given by
\bea
ds_{10\rm{IIA}}^2 &=& H^{-5/8} K^{-1/8}\, (-dt^2 + dw_1^2 + dw_2^2) +
H^{3/8} K^{7/8}\, (dz_1^2 + dz_2^2 + dz_3^2)\nn\\
&&+H^{3/8} K^{-1/8}\, ( \wtd K\,(dx^2 + x^2\, d\wtd \Omega_2^2) +
\wtd K^{-1} (dy + \wtd Q_{\sst{\rm N}}\, \wtd \omega)^2)\,,\nn\\
e^{\phi} &=& H^{1/4} K^{-3/4}\,,\label{d2d6nut}\\
F_\4&=&dt\wedge d^2w\wedge dH^{-1}\,,\qquad
F_2=e^{-3/2\phi} {*(dt\wedge d^2w\wedge d^4x\wedge dK^{-1})}\,.\nn
\eea
where $x^2 = x_1^2+ x_2^2 + x_3^2$ and the functions $H$, $K$ and
$\wtd K$ are given by
\be
H=1 + \fft{Q}{(4\wtd Q_{\sst{\rm N}}\, x + 4 \Qn\, z)^3}\,,\quad
K=\fft{\Qn}{z}\,,\quad \wtd K= \fft{\wtd Q_{\sst{\rm N}}}{x}\,.
\ee
It is instructive to illustrate the solution in the following diagram:

\bigskip\bigskip
\centerline{
\begin{tabular}{c|ccccccccccc}
&$t$ & $w_1$ & $w_2$ & $x_1$ & $x_2$ & $x_3$ & $y$ & $z_1$ & $z_2$ &
$z_3$ &\\ \hline
D2&$\times$ & $\times$ & $\times$ & $-$ &$-$ &$-$ &$-$ &$-$ &$-$
&$-$ & $H$ \\
D6&$\times$ & $\times$ & $\times$ & $\times$ & $\times$ & $\times$
&$\times$ &$-$ &$-$ &$-$ & $K$ \\
NUT& $\times$ & $\times$ & $\times$ & $-$ & $-$ & $-$ & $*$ &
$\times$ & $\times$ & $\times$ & $\wtd K$
\end{tabular}}
\bigskip

\centerline{Diagram 9. The D2/D6/NUT system}
\bigskip\bigskip

We can now perform the T-duality on the coordinate $y$, and obtain the
semi-localised D3/D5/NS5 intersection of the type IIB theory, given by
\bea
ds_{10\rm{IIB}}^2 &=& H^{-1/2}(K\,\wtd K)^{-1/4}\, \Big[
-dt^2 + dw_1^2 + dw_2^2\nn\\
&& H\, \wtd K\, (dx_1^2 + dx_2^2 + dx_3^2) + 
K\, \wtd K\, dy^2 + H\, K\, (dz_1^2 + dz_2^2 + dz_3^2)\Big]\,.
\eea
It is straightforward to verify that the near-horizon structure of the above
D3/D5/NS5 system is of the form (\ref{d3d5ns5horizon}).   The solution
can be illustrated by the following diagram:

\newpage
\centerline{
\begin{tabular}{c|ccccccccccc}
&$t$ & $w_1$ & $w_2$ & $x_1$ & $x_2$ & $x_3$ & $y$ & $z_1$ & $z_2$ &
$z_3$ &\\ \hline
D3&$\times$ & $\times$ & $\times$ & $-$ &$-$ &$-$ &$\times$ &$-$ &$-$
&$-$ & $H$ \\
D5&$\times$ & $\times$ & $\times$ & $\times$ & $\times$ & $\times$
&$-$ &$-$ &$-$ &$-$ & $K$ \\
NS5& $\times$ & $\times$ & $\times$ & $-$ & $-$ & $-$ & $-$ &
$\times$ & $\times$ & $\times$ & $\wtd K$
\end{tabular}}
\bigskip

\centerline{Diagram 10. The D3/D5/NS5 system}
\bigskip\bigskip

\subsection{M2/pp-wave system}

          The M2/pp-wave solution is given by
\bea
ds_{11}^2 &=& H^{-2/3}(-W^{-1}\, dt + W\, (dy + (W^{-1}-1)dt)^2 +
dx^2) + H^{1/3}(dz^2 + z^2\, d\Omega_7^2)\,,\nn\\
F_\4 &=& dt\wedge dy\wedge dx\wedge dH^{-1}\,.\label{m2wave}
\eea
The solution can be illustrated by the following diagram:

\bigskip\bigskip
\centerline{
\begin{tabular}{c|cccccccccccc}
&$t$ & $y_1$ & $x_1$ & $z_1$ & $z_2$ & $z_3$ & $z_4$ & $z_5$ & $z_6$ &
$z_7$ & $z_8$ & \\ \hline
M2&$\times$ & $\times$ & $\times$ & $-$ &$-$ &$-$ &$-$ &$-$ &$-$
&$-$ &$-$ & $H$ \\
wave&$\times$ & $\sim$ & $-$ & $-$ & $-$ & $-$
&$-$ &$-$ &$-$ &$-$ & $-$ & $W$ \\
\end{tabular}}
\bigskip

\centerline{Diagram 11. The M2/pp-wave brane intersection.}
\bigskip\bigskip

   When both functions $H$ and $W$ are harmonic on the overall
transverse space of the $z^i$ coordinates, the metric becomes a direct
product of the Kaigorodov metric with a 7-sphere in the near-horizon
limit.  Here, we instead consider a semi-localised solution, with $H$
and $K$ given by
\be
H=\fft{Q}{z^6}\,,\qquad W= 1 + Q_w\, (x^2 + \fft{Q/4}{z^4})\,.
\ee
Making the coordinate transformation
\be
x=\fft{\cos\a}{r}\,,\qquad z^2 = \fft{r\, Q^{1/2}}{2\sin\a}\,,
\ee
the metric becomes AdS$_4\times S^7$, with
\be
ds_{11}^2 = \fft{Q^{1/3}}{4s^2}(ds_{\rm{AdS}_3}^2 + d\a^2)+
 Q^{1/3}\, d\Omega_7^2\,.\label{m2wavehorizon}
\ee
Here $ds_{\rm{AdS}_3}^2$ is the metric of AdS$_3$ (the BTZ black
hole), given by (\ref{ads3}).  Thus, we have demonstrated that the
semi-localised M2/pp-wave system is a warped product of AdS$_3$ and an
8-space.  Making the coordinate transformation $\tan(\a/2) = e^\rho$,
the first part of (\ref{m2wavehorizon}) can be expressed as
\be
ds_4^2 = d\rho^2 + \cosh^2\rho\, ds_{\rm{AdS}_3}^2\,.
\ee
This is AdS$_4$ expressed as a foliation of AdS$_3$ (see appendix A).

\subsection{The NS1/D0 system}

         Reducing the above solution on the coordinate $y_1$, it 
becomes an intersecting NS1/D0 system, with
\bea
ds_{10\rm{IIA}} &=& H^{-3/4}W^{-7/8}\Big(-dt^2 + W\, dx^2 + W\,H\,(dz_1^2 + 
\cdots + dz_8^2)\Big)\,,\nn\\
F_\3&=& dt\wedge dx\wedge dH^{-1}\,,\qquad
F_\2= dt\wedge dW^{-1}\,,\nn\\
e^{\phi} &=& H^{-1/2}\, W^{3/4}\,.\label{ns1d0}
\eea
The metric of the near-horizon region describes a warped product
of AdS$_2$ with an 8-space:
\be
ds_{10\rm{IIA}}^2 = 8^{-3/4}Q^{3/8}Q_w^{1/8}\, s^{-9/4}\, 
(ds_{\rm{AdS}_2}^2 + d\a^2 +4s^2 d\Omega_7^2)\,,
\ee
where $ds_{\rm{AdS}_2}^2$ is the metric of AdS$_2$, given by
(\ref{ads2}).  The NS1/D0 system can be illustrated by the following
diagram:

\bigskip\bigskip
\centerline{
\begin{tabular}{c|cccccccccccc}
&$t$ & $x_1$ & $z_1$ & $z_2$ & $z_3$ & $z_4$ & $z_5$ & $z_6$ &
$z_7$ & $z_8$ & \\ \hline
NS1&$\times$ & $\times$ & $-$ &$-$ &$-$ &$-$ &$-$ &$-$
&$-$ &$-$ & $H$ \\
D0&$\times$ & $-$ & $-$ & $-$ & $-$
&$-$ &$-$ &$-$ &$-$ & $-$ & $W$ \\
\end{tabular}}
\bigskip

\centerline{Diagram 12.  The NS1/D0 brane intersection.}
\bigskip\bigskip

    In the M2/pp-wave and NS1/D0 systems, the internal space has a
round 7-sphere.  We can replace it by foliating of two lens spaces
$S^3/Z_{\Qn}$ and $S^3/Z_{\wtd Q_{\sst{\rm N}}}$.  As discussed in the
appendix B, this can be achieved by introducing two NUTs in the
intersecting system.  We can then perform Kaluza-Klein reductions or
T-duality transformations on the two associated fibre coordinates of
the lens spaces.  The resulting configurations can then be viewed as
the near-horizon geometries of four intersecting $p$-branes, with
charges $Q$, $Q_w$, $\Qn$ and $\wtd Q_{\sst{\rm N}}$

\section{M5/NUT and M5/pp-wave systems}

\subsection{M5/NUT and NS5/D6 systems}

      The solution of an M5-brane intersecting with a NUT is given by
\bea
ds_{11}^2\!\!\! &=&\!\!\! H^{-1/3}(-dt^2 + dw_1^2 + \cdots +dw_5^2) +
H^{2/3}(dx_1^2 + K\, (dz^2 + z^2 d\Omega_2^2) +
K^{-1}(dy + \omega)^2)\,,\nn\\
F_\4 &=& {*(dt\wedge d^5w\wedge dH^{-1})}\,.\label{m5nut}
\eea
The solution can be illustrated by the following diagram:

\bigskip\bigskip
\centerline{
\begin{tabular}{c|cccccccccccc}
&$t$ & $w_1$ & $w_2$ & $w_3$ & $w_4$ & $w_5$ & $x_1$ & $z_1$ & $z_2$ &
$z_3$ & $y$ & \\ \hline
M5&$\times$ & $\times$ & $\times$ & $\times$ &$\times$ &$\times$ 
&$-$ &$-$ &$-$ &$-$ &$-$ & $H$ \\
NUT&$\times$ & $\times$ & $\times$ & $\times$ & $\times$ & $\times$
&$\times$ &$-$ &$-$ &$-$ &$*$ & $K$ \\
\end{tabular}}
\bigskip

\centerline{Diagram 13.  The M5/NUT brane intersection.}
\bigskip\bigskip

In the usual construction where the harmonic functions $H$ and $K$
depend only the $z$ coordinate, the metric does not have an AdS
structure in the near-horizon region.  Here, we instead consider a
semi-localised solution, given by
\be
H=1 + \fft{Q}{(x^2 + 4\Qn\, z)^{3/2}}\,,\qquad
K = \fft{\Qn}{z}\,.
\ee
After an analogous coordinate transformation, we find that the metric
can be expressed as
\bea
ds_{11}^2 &=& H^{-1/3}(-dt^2 + dw_i\, dw_i) + H^{2/3}(dr^2+ r^2\,
dM_4^2)\,,\nn\\
dM_4^2 &=& d\a^2 + \ft14 s^2\, (d\Omega_2^2 + (\fft{dy}{\Qn} + 
\omega)^2)\,.
\eea
Thus, in the near-horizon limit, the metric is AdS$_7\times M_4$, where
$M_4$ is a foliation of a lens space $S^3/Z_{\Qn}$.

          We can dimensionally reduce the solution (\ref{m5nut}) on
the fibre coordinate $y$. The resulting solution is the NS-NS 5-brane
intersecting with a D6-brane:

\bigskip\bigskip
\centerline{
\begin{tabular}{c|ccccccccccc}
&$t$ & $w_1$ & $w_2$ & $w_3$ & $w_4$ & $w_5$ & $x_1$ & $z_1$ & $z_2$ &
$z_3$ & \\ \hline
NS5&$\times$ & $\times$ & $\times$ & $\times$ &$\times$ &$\times$ 
&$-$ &$-$ &$-$ &$-$ & $H$ \\
D6&$\times$ & $\times$ & $\times$ & $\times$ & $\times$ & $\times$
&$\times$ &$-$ &$-$ &$-$ & $K$ \\
\end{tabular}}
\bigskip

\centerline{Diagram 14. The NS5/D6 brane intersection.}
\bigskip\bigskip

The solution is given by
\bea
ds_{10\rm{IIA}}^2 &=& H^{-1/4}K^{-1/8}\, (-dt^2 + dw_i\, dw_i) + H^{3/4}
K^{-1/8}\, dx^2 + H^{3/4} K^{7/8}\, dz_i\, dz_i\,,\nn\\
e^{\phi} &=& H^{1/2} K^{-3/4}\,,\qquad
F_\3 = e^{\phi/2} {*(dt\wedge d^5w\wedge dH^{-1})}\,,\nn\\
F_\2 &=& e^{-3\phi/2} {*(dt\wedge d^5w\wedge dx\wedge dK^{-1})}\,,
\eea 
In the near-horizon limit, the metric becomes a warped product of
AdS$_7$ with a 3-space
\be
ds_{10\rm{IIA}}^2 = \fft{Q^{3/4}}{(2\Qn)^{1/4}}\,s^{1/4}\, 
(\fft{r}{Q}(-dt^2 + dw_i\, dw_i) + \fft{dr^2}{r^2} + d\a^2
+ \ft14 s^2\, d\Omega_2^2)\,.
\ee

\subsection{M5/pp-wave and D0/D4 system}

       The solution of an M5-brane with a pp-wave is given by
\bea
ds_{11}^2 &=& H^{-1/3}(-W^{-1}\, dt^2 + W\, (dy_1 + (W^{-1} -1)dt)^2 +
dx_1^2 + \cdots + dx_4^2)\nn\\
&&\phantom{xxxxxxxxxxxxx} + H^{2/3}\, 
(dz_1^2 + \cdots + dz_5^2)\,,\nn\\
F_4&=&{*(dt\wedge dy_1\wedge d^4x\wedge dH^{-1})}\,.\label{m5wave} 
\eea
The solution can be illustrated by the following diagram:

\bigskip\bigskip
\centerline{
\begin{tabular}{c|cccccccccccc}
&$t$ & $y_1$ & $x_1$ & $x_2$ & $x_3$ & $x_4$ & $z_1$ & $z_2$ & $z_3$ &
$z_4$ & $z_5$ & \\ \hline
M5&$\times$ & $\times$ & $\times$ & $\times$ &$\times$ &$\times$ 
&$-$ &$-$ &$-$&$-$ &$-$ & $H$ \\
wave&$\times$ & $\sim$ & $-$ & $-$ & $-$ & $-$
&$-$ &$-$ &$-$ &$-$ & $-$ & $W$ \\
\end{tabular}}
\bigskip

\centerline{Diagram 15. The M5/pp-wave brane intersection.}
\bigskip\bigskip

  We shall consider semi-localised solutions, with the functions $H$
and $W$ given by
\be
H=\fft{Q}{z^3}\,,\qquad W=1 + Q_w\, (x^2 + \fft{4Q}{z})\,.
\ee
Using analogous coordinate transformations, we find that the metric
of the semi-localised M5/pp-wave system becomes
\be
ds_{11}^2 = 4Q^{2/3}\, s^{-2} (ds_{\rm AdS_3}^2 + d\a^2 + c^2\, 
d\Omega_3^2) + Q^{2/3}\, d\Omega_4^2\,,\label{m5wavehorizon}
\ee
where $ds_{\rm{AdS}_3}^2$, given by (\ref{ads3}), is precisely the
extremal BTZ black hole and hence is is locally AdS$_3$. After making
the coordinate transformation $\tan(\a/2) = e^\rho$, the first part of
the metric (\ref{m5wavehorizon}) can be expressed as
\be
ds_7^2 = d\rho^2 + \sinh^2\rho\, d\Omega_3^2 + 
\cosh^2\rho\, ds_3^2\,.
\ee
This is AdS$_7$ written as a foliation of AdS$_3$ and $S^3$.

         Performing a dimensional reduction of the solution
(\ref{m5wave}) on the coordinate $y_1$, we obtain a D0/D4 intersecting
system, given by
\bea
ds_{10\rm{IIA}}^2 &=& H^{-3/8} W^{-7/8}(-dt^2 + W\, dx_i\, dx_i +
H\, W\, dz_i\, dz_i)\,,\nn\\
e^{\phi} &=& H^{-1/4}\, W^{3/4}\,,\qquad
F_\2=dt\wedge dW^{-1}\,,\nn\\
F_4&=& e^{-\phi/2} {*(dt\wedge d^4x\wedge dH^{-1})}\,.\label{d0d4}
\eea
The near-horizon limit of the semi-localised D0/D4 system is a
warped product of AdS$_2$ with an 8-space:
\be
ds_{10\rm{IIA}}^2= 2^{9/4}Q^{3/4}\, Q_w^{1/8}\, s^{-9/4}
\, (ds_2^2 + d\a^2 + c^2\, d\Omega_3^2 + \ft14 s^2\, d\Omega_4^2)\,, 
\ee
where $ds_2^2$ is given by (\ref{ads2}).  We illustrate this
intersecting system with the following diagram

\bigskip\bigskip
\centerline{
\begin{tabular}{c|ccccccccccc}
&$t$ & $x_1$ & $x_2$ & $x_3$ & $x_4$ & $z_1$ & $z_2$ & $z_3$ &
$z_4$ & $z_5$ & \\ \hline
D4&$\times$ & $\times$ & $\times$ &$\times$ &$\times$ 
&$-$ &$-$ &$-$&$-$ &$-$ & $H$ \\
D0&$\times$ & $-$ & $-$ & $-$ & $-$
&$-$ &$-$ &$-$ &$-$ & $-$ & $W$ \\
\end{tabular}}
\bigskip

\centerline{Diagram 16. The D0/D4 brane intersection.}
\bigskip\bigskip

        In this example in the internal space the round $S^3$ and
$S^4$ can be replaced by a lens space $S^3/Z_{\Qn}$ and the foliation of
a lens space $S^3/Z_{\wtd Q_{\sst{\rm N}}}$, respectively.  We can
then perform Kaluza-Klein reductions or T-duality transformations on
the fibre coordinates of the lens spaces, leading to four-component
intersections with charges $Q$, $Q_w$, $\Qn$ and $\wtd Q_{\sst{\rm
N}}$.

\section{AdS$_6$ in type IIB from T-duality}

       So far in this paper we have two examples of intersecting
D$p$/D$(p+4)$ systems in the type IIA theory that give rise to warped
products of AdS$_{p+2}$ with certain internal spaces, namely for $p=0$
and $p=2$.  It was observed \cite{bo} also that the D4/D8 system,
arising from massive type IIA supergravity, gives rise to the warped
product of AdS$_6$ with a 4-sphere in the near-horizon limit:
\be
ds_{10\rm{IIA}}^2 = s^{1/12}\, (ds_{\rm AdS_6}^2 + g^{-2}(d\a^2 + c^2\,
d\Omega_3^2)) \,.\label{ads6type2a}
\ee
Note that the D4/D8 system is less trivial than the previous examples,
in the sense that it cannot be mapped by T-duality to a non-dilatonic
$p$-brane intersecting with a NUT or a wave.

       We can now introduce a NUT in the intersecting system which
has the effect, in the near-horizon limit, of replacing the round
3-sphere by a lens space, given in (\ref{lens}).  We can then perform
a Hopf T-duality transformation and obtain an embedding of AdS$_6$ in
type IIB theory:
\be
ds_{10}^2 = c^{1/2}\, \Big[ds_{\rm{AdS}_6}^2 +
g^{-2}(d\a^2 + \ft14 c^2\, d\Omega_2^2) +  s^{2/3}\, c^{-2}\,
dy^2\Big]\,.\label{ads6type2b}
\ee
This solution can be viewed as the near-horizon geometry of an
intersecting D5/D7/NS5 system.  It provides a background for the exact
embedding of six-dimensional gauged supergravity in type IIB theory.

         The D5/D7/NS5 semi-localised solution can be obtained by
performing the T-duality on the D4/D8/NUT system.  The solution is
given by
\bea
ds_{10\rm{IIB}}^2 &=& (H_1\, K)^{-1/4}\Big(-dt^2 + dw_1^2 + \cdots +
dw_4^2 + H_1\, K\, (dx_1^2 + dx_2^2 + dx_3^2)\nn\\
&&\phantom{xxxxxx} + H_2\, K\, dy^2 + H_1\, H_2\, dz^2\Big)\,.
\eea
The functions $H_1$, $H_2$ and $K$ are given by
\be
H_1 =1+ \fft{Q_1}{(4\Qn\, |\vec x| + \ft{4Q_2}{9}\, z^3)^{5/3}}\,,\quad
H_2=Q_2\, z\,,\quad K = \fft{\Qn}{|\vec x|}\,.
\ee
It is straightforward to verify that the near-horizon structure of
this system is of the form (\ref{ads6type2b}). The solution can be
illustrated by the following:

\bigskip\bigskip
\centerline{
\begin{tabular}{c|ccccccccccc}
&$t$ & $w_1$ & $w_2$ & $w_3$ & $w_4$ & $x_1$ & $x_2$ & $x_3$ &
$y$ & $z$ & \\ \hline
D5&$\times$ & $\times$ & $\times$ &$\times$ &$\times$ 
&$-$ &$-$ &$-$&$-$ &$-$ & $H_1$ \\
D7&$\times$ & $\times$ & $\times$ & $\times$ & $\times$
&$\times$ &$\times$ &$\times$ &$-$ & $-$ & $H_2$ \\
NS5& $\times$ & $\times$ & $\times$ & $\times$ & $\times$ &
$-$ & $-$ & $-$ & $-$ & $\times$ & $K$ \\
\end{tabular}}
\bigskip

\centerline{Diagram 17. The D5/D7/NS5 brane intersection.}
\bigskip\bigskip

\section{Conclusion}

          In this paper, we obtain various AdS spacetimes warped with
certain internal spaces in eleven-dimensional and type IIA/IIB
supergravities.  These solutions arise as the near-horizon geometries
of more general semi-localised multi-intersections of M-branes in
$D=11$ or NS-NS branes or D-branes in $D=10$.  We achieve this by
noting that any bigger sphere (AdS spacetime) can be viewed as a
foliation involving $S^3$ (AdS$_3$).  Then the $S^3$ (AdS$_3$) can be
replaced by a three-dimensional lens space (BTZ black hole), which
arise naturally from the introduction of a NUT (pp-wave). We can
then perform a Kaluza-Klein reduction or Hopf T-duality transformation
on the fibre coordinate of the lens space (BTZ black hole).

          It is important to note that the warp factor depends only on
the internal foliation coordinate but not on the lower-dimensional
spacetime coordinates.  This implies the possibility of finding a
larger class of consistent dimensional reduction of eleven-dimensional
or type IIA/IIB supergravity on the internal space, giving rise to
gauged supergravities in lower dimensions with AdS vacuum solutions.
The first such example was obtained in \cite{d6gauge}.  In this paper,
we obtain further examples for possible consistent embeddings of
lower-dimensional gauged supergravity in $D=11$ and $D=10$.  For
example, we obtain the vacuum solutions for the embedding of the six
and four-dimensional gauged AdS supergravities in type IIB theory and
for the embedding of the seven-dimensional gauged AdS supergravity in
type IIA theory.

\section*{Acknowledgement}

       C.N.P. would like to thank the Caltech-USC Center for
Theoretical Physics, and Imperial College, London,  for hospitality
during the course of this work.

\appendix

\section{Spheres and AdS from foliations}

   There are two closely parallel constructions which arise in the
various intersections involving NUTs and waves.  The former involves a
construction of the unit metric on the sphere $S^{p+q+1}$ as a
foliation of $S^p\times S^q$ surfaces, while the latter involves an
analogous construction of the unit metric on AdS$_{p+q+1}$, as a
foliation of AdS$_p \times S^q$ surfaces.

   Consider first the construction of the unit $S^{p+q+1}$ metric.  We
start from the unit metrics $d\Omega_p^2=dX^i\, dX^i$ and
$d\Omega_q^2= dY^a\, dY^a$ on the spheres $S^p$ and $S^q$, defined as
the surfaces
\be
X^i\, X^i =1\,,\qquad Y^a\, Y^a = 1
\ee
in $\R^{p+1}$ and $\R^{q+1}$ respectively.  We now introduce Cartesian
coordinates $Z^A=(Z^i,Z^a)$ in $\R^{p+q+2}$, defined by
\be
Z^i = X^i\, \cos\a\,,\qquad Z^a = Y^a\, \sin\a\,,\label{spq2}
\ee
and so $Z^A\, Z^A=1$, thus defining a unit sphere $S^{p+q+1}$ in
$\R^{p+q+2}$.  Clearly (\ref{spq2}) defines a complete
parameterisation of points in $\R^{p+q+2}$, with $0\le\a\le\ft12 \pi$,
and so $\a$, together with the constrained coordinates $x^i$ and $y^a$
on the spheres $S^p$ and $S^q$, provide coordinates for the unit
sphere $S^{p+q+1}$ with a manifest $SO(p+q+2)$ isometry group action
on the $Z^A$ coordinates.  The metric on $S^{p+q+1}$ is given by
$d\Omega_{p+q+1}^2 =dZ^A\, dZ^A$, and so from the above definitions we
obtain
\be
d\Omega_{p+q+1}^2 = d\a^2 + \cos^2\a\, d\Omega_p^2 + \sin^2\a\,
d\Omega_q^2\,.
\ee
The foliating surfaces at a fixed value of the ``latitude''
coordinate $\a$ are $S^p\times S^q$, with radii
$\cos\a$ and $\sin\a$ for the two factors.  The construction is a
generalisation of the Clifford Torus $S^1\times S^1$ foliating $S^3$.

   In a similar manner, one can construct a metric $d\omega_{p+q+1}^2$
on the unit AdS$_{p+q+1}$ as follows.  We start from a unit AdS$_p$,
with metric $d\omega_p^2= dX^\mu\, dX^\nu\, \eta_{\mu\nu}$, and a unit
$S^q$ with metric $d\Omega_q^2 = dY^a\, dY^a$, where the coordinates
$X^\mu$ on $\R^{p+1}$ satisfy the indefinite-signature condition
\be
X^\mu\, X^\nu \, \eta_{\mu\nu}=-1\,,\qquad \eta_{\mu\nu}
=\hbox{diag}(-1,-1,1,1,\ldots, 1)\,,
\ee
while the coordinates $Y^a$ on $\R^{q+1}$ satisfy $Y^a\, Y^a=1$ as before.
We now define coordinates $Z^A=(Z^\mu,Z^a)$ by
\be
Z^\mu = X^\mu\, \cosh\rho \,,\qquad Z^a = Y^a\, \sinh\rho\,,
\ee
which therefore satisfy
\be
Z^A\, Z^B\, \eta_{AB}=-1\,,\qquad \eta_{AB}=
\hbox{diag}(-1,-1,1,1,\ldots, 1)\,.
\ee
The coordinates $Z^A$, subject to this constraint, therefore define
AdS$_{p+q+1}$, with a manifest $SO(p+q-1,2)$ isometry.  The metric
$d\omega_{p+q+1}^2 = dZ^A\, dZ^B\, \eta_{AB}$ is given by
\be
d\omega_{p+q+1}^2 = d\rho^2 + \cosh^2\rho\, d\omega_{p}^2 +
\sinh^2\rho\, d\Omega_q^2\,.
\ee

\section{NUTs without NUTs}

    In this appendix, we show explicitly that the semi-localised
intersection of a $p$-brane with a Kaluza-Klein monopole (a NUT) can
be recast, after appropriate coordinate transformations, as a
restricted class of ordinary distributed $p$-branes.  For
definiteness, we take the case of a semi-localised intersection of the
M2-brane with a NUT as an example.  The analysis for the other cases
is essentially identical.
 
   The semi-localised solution obtained in \cite{youm} is given by
\bea
ds_{11}^2 &=& H^{-2/3}\, dw^\mu\, dw_\mu + H^{1/3}\,[ (dx_1^2 +
\cdots + dx_4^2)  \nn\\
&&\qquad \qquad + K\, (dz_1^2 + dz_2^2 + dz_3^2) + K^{-1}\, (dy +
A_i\, dz_i)^2]\,,\nn\\
K&=& \fft{\Qn}{|\vec z|} \,,\qquad A_i\, dz_i = \Qn\, \cos\theta\,
d\varphi\,,\label{m2kk}\\
H &=& 1+ \sum_k \fft{Q_k}{\Big( |\vec x-\vec x_{0k}|^2 + 4 \Qn\, |\vec
z|\Big)^3}\,,\nn
\eea
where $Q_k$ denotes the M2-brane charge located at $\vec x_{0 k}$, 
$\Qn$ is the NUT charge, and we take 
\be
(z_1,z_2,z_3) = \fft{R^2}{4 \Qn}\, (\sin\theta\, \cos\varphi,
\sin\theta\, \sin\varphi,\cos\theta)\,.
\ee

   It now follows that the part of the metric
\be
d\bar s^2 \equiv K\, (dz_1^2 + dz_2^2 + dz_3^2) + K^{-1}\, (dy +
A_i\, dz_i)^2\label{dsbar}
\ee
is nothing but the locally-flat metric 
\be
d\bar s^2 = dR^2 + R^2\, d\bOmega_3^2\,,
\ee
where
\be
d\bOmega_3^2 \equiv \ft14 d\Omega_2^2 + \ft14
\Big(\fft{dy}{\Qn} +\cos\theta\, d\varphi\Big)^2
\ee
is the metric on the cyclic lens space $S^3/Z_{\Qn}$.  Locally, this
is just the standard metric on the unit 3-sphere.  Viewed as a $U(1)$
bundle over $S^2$ the coordinate $y$ on the $U(1)$ fibres is taken
always to have the period $4\pi$.  When $\Qn=1$, the topology is
therefore precisely $S^3$.  However, if $\Qn$ is a larger integer, the
fibre coordinate has a period that is smaller by the fraction $1/\Qn$
than the period that would be needed for $S^3$ itself, and
consequently the topology is $S^3/Z_{\Qn}$.

    The solution (\ref{m2kk}) can therefore be recast as
\be
ds_{11}^2 = H_2^{-2/3}\, dw^\mu\, dw_\mu + H_2^{1/3}\, (dx_1^2 +
\cdots + dx_4^2 + d\td z_1^2 + \cdots + d\td z_4^2)\,,\label{m2dist}
\ee
with the harmonic function given by
\be
H_2 = 1 + \sum_k \fft{Q_k}{\Big( |\vec x-\vec x_{0k}|^2 
+ |\vec {\td z}|^2 \Big)^3}\,.\label{m2dist2}
\ee
The coordinates $\td z_i$ live on $\R^4/Z_{\Qn}$, and are related to
$R$ and the coordinates $(\theta,\varphi, y)$ on the lens space
$S^3/Z_{\Qn}$ by
\be
\td z_1 + \im\, \td z_2 = R\, \sin\ft12\theta\, e^{\fft{\im}{2}(y/\Qn
+ \varphi)}\,,\qquad
\td z_3 + \im\, \td z_4 = R\, \cos\ft12\theta\, e^{\fft{\im}{2}(y/\Qn
- \varphi)}\,.
\ee
In other words, if we make the following coordinate transformation from
$(z_1,z_2,z_3,y)$ to $(\td z_1, \td z_2, \td z_3, \td z_4)$,
\bea
\td z_1 + \im\, \td z_2 &=& \Big[
\fft{2 \Qn\, (r+z_3)(z_1 +\im\, z_2)}{\sqrt{z_1^2 +
z_2^2}}\Big]^{1/2}\, e^{\fft{\im}{2\Qn}\, y}\,,\nn\\
\td z_3 + \im\, \td z_4 &=& \Big[
\fft{2 \Qn\, (r-z_3)(z_1 -\im\, z_2)}{\sqrt{z_1^2 +
z_2^2}}\Big]^{1/2}\, e^{\fft{\im}{2\Qn}\, y}\,,
\eea
where $r^2\equiv z_1^2 + z_2^2 + z_3^2$, then the metric (\ref{dsbar})
is seen to be nothing but
\be
d\bar s^2 = d\td z_1^2 + d\td z_2^2 + d\td z_3^2 + d\td z_4^2\,.
\label{dsbar2}
\ee

   The semi-localised M2-brane/NUT intersection (\ref{m2kk}) can
therefore be obtained by starting from a standard distribution of pure
M2-branes (\ref{m2dist}), with charges spread over only four of the
eight transverse directions as in (\ref{m2dist2}).  This is precisely
equivalent to the semi-localised M2-brane/NUT intersection
(\ref{m2kk}) with unit NUT charge, $\Qn=1$.  To obtain higher values
of the NUT charge, one simply has to factor the $\R^4$ space of the
$\td z_i$ coordinates by $Z_{\Qn}$, as defined above.  Note that
although this semi-localised way of introducing a NUT seems trivial,
in that it amounts a coordinate transformation, performing
Kaluza-Klein reduction on the fibre coordinate does create a
non-trivial intersecting component, since the Kaluza-Klein 2-form
field strength now carries a non-trivial flux.

    The above discussion carries over, {\it mutatis mutandis},  to the
cases of the semi-localised M5-brane/NUT and D3-brane/NUT.

\end{document}